\documentclass[aps,prl,twocolumn,showpacs,preprintnumbers]{revtex4-1}

\usepackage{psfrag,graphicx}
\usepackage{dcolumn}
\usepackage{amsmath,amssymb}
\usepackage{bm}
\usepackage{amsfonts,amssymb,amsmath}        % for math symbols.
\usepackage{epstopdf}

\newcommand{\be}{\begin{equation}}
\newcommand{\ee}{\end{equation}}
\newcommand{\bq}{\begin{eqnarray}}
\newcommand{\eq}{\end{eqnarray}}

\newcommand{\no}{\nonumber\\}

\newcommand{\ket}[1]{\left | \, #1 \right\rangle}

\begin{document}

\title{Bringing order through disorder: Localisation of errors in topological quantum memories}
\date{\today}
\author{James R. Wootton and Jiannis K. Pachos}
\affiliation{School of Physics and Astronomy, University of Leeds, Leeds LS2 9JT, U.K.}

\begin{abstract}

%The ability to reliably store quantum states is an essential element for any task in quantum information. Topological systems promise to protect quantum information by topological and energetic considerations. If undesired anyonic excitations are propagated at large distances, either through coherent or probabilistic processes, they can cause logical errors in the topologically encoded information. Here we show that Anderson localisation induced by disorder in the system can successfully protect topological quantum memories from the coherent propagation of anyons. For concreteness we employ the toric code model at zero temperature. It is known that in the absence of a magnetic field it can tolerate a finite initial density of anyonic errors. In the presence of a spurious magnetic field anyonic quantum walks are induced and the tolerable density becomes zero. We demonstrate that disorder in the couplings of the model can successfully localise anyons. This allows the topological quantum memory to tolerate a finite initial density of anyonic errors for arbitrarily long times. We anticipate that disorder inherent in any physical realisation of topological systems will help to strengthen the fault-tolerance of quantum memories.

Anderson localization emerges in quantum systems when randomised parameters cause the exponential suppression of motion. Here we consider this phenomenon in topological models and establish its usefulness for protecting topologically encoded quantum information. For concreteness we employ the toric code. It is known that in the absence of a magnetic field this can tolerate a finite initial density of anyonic errors, but in the presence of a field anyonic quantum walks are induced and the tolerable density becomes zero. However, if the disorder inherent in the code is taken into account, we demonstrate that the induced localization allows the topological quantum memory to regain a finite critical anyon density, and the memory to remain stable for arbitrarily long times. We anticipate that disorder inherent in any physical realisation of topological systems will help to strengthen the fault-tolerance of quantum memories.

\end{abstract}


\maketitle

{\bf Introduction:} {Topological quantum memories are many-body interacting systems that can serve as error correcting codes \cite{kitaev}. These models possess degenerate ground state manifolds in which quantum information may be encoded. The size of the model and its energy gap then protects this information, preventing local perturbations from splitting the degeneracy and hence causing errors \cite{trebst,jiang,vidal,hastings}. However, the dynamic effects of perturbations when excitations are present are a serious problem for the stability of the memory \cite{kay}, especially for non-zero temperature \cite{Chamon}. Several promising schemes have been proposed \cite{castelnovo,dimitris} that suggest ways to combat this problem with their own merits and drawbacks. In particular, in \cite{dimitris} it is shown that disorder can aid the stability of topological phases. Here we shed new light on the issue by showing that disorder in topological memories can induce Anderson localization \cite{anderson}. This exponentially suppresses the dynamic effects of perturbations on exited states, and allows the memory to remain stable.}

%{\bf Introduction:} Topological quantum memories are error correcting codes converted into many-body interacting systems \cite{kitaev}. It has been shown that non-locality and an energy gap can protect the topologically encoded information against local perturbations \cite{trebst,jiang,vidal,hastings}. Nevertheless, the dynamic action of these perturbations \cite{kay} as well as a non-zero temperature \cite{Chamon,dennis} can lead to the memory becoming unstable. Several promising schemes have been proposed \cite{castelnovo,chesi,dimitris} that suggest ways to combat this problem with their own merits and drawbacks. In particular, in \cite{dimitris} it is shown that disorder can aid the stability of topological phases. Here we shed new light on this issue by considering the stability of topological quantum memories against dynamic effects at zero temperature. We demonstrate that Anderson localisation can neutralise the environmental perturbations when the system is subject to disorder by exponentially suppressing the propagation of errors \cite{anderson,lewenstein}.

In the definition of a topological memory, we require the following conditions. First, the stored information is encoded within a degenerate ground state of a system. Second, we require that the memory can be left exposed to some assumed noise model for arbitrarily long times without active monitoring or manipulations. Finally, measurement of the system after errors have occurred extracts both the (now noisy) contents of the memory and an error syndrome, allowing a one-off error correction step to be performed to retrieve the original stored information.

In topological memories, errors create anyonic excitations. Logical errors correspond to the propagation of these anyons around topologically non-trivial paths on the surface of the system. While the anyons are normally static, a kinetic term emerges in the presence of a spurious magnetic field \cite{kay}. If this can act unchecked, even a single pair of anyons will cause a logical error after a time linear with the system size, $L$. This means that the memory is not resilient against any non-zero density of anyons initially present in the system. We demonstrate that randomness in the couplings of the code cause anyons to remain well localised in their initial positions. This enables the topological memory to successfully store quantum information for arbitrarily long times as long as the distribution of anyons is below a critical value. Hence, topological quantum memories can be made fault-tolerant against the dynamical effects of local perturbations.

{\bf The toric code:} The toric code is the simplest topological memory. It is a quantum error correcting code whose code distance depends explicitly on the linear size of the system \cite{kitaev}. The code can be defined on a two-dimensional $L \times L$ square lattice wrapped around a torus with a spin-$1/2$ particle at each edge. Stabiliser operators are defined on the spins around each vertex, $v$ and plaquette, $p$, of the lattice, $A_v = \prod_{i \in v} \sigma^x_i$, $B_p = \prod_{i \in p} \sigma^z_i$. The stabilizer space, where information is stored, is composed of states which satisfy $A_v \ket{\xi} = \ket{\xi}$ and $B_p \ket{\xi} = \ket{\xi}$ for all $v$ and $p$. Violations of these stabilizers on plaquettes or vertices are associated with quasiparticles known as anyons, with so-called $e$ anyons on the vertices and $m$ anyons on plaquettes \cite{kitaev}. {These quasiparticles are hardcore bosons, though they also have non-trivial anyonic statistics with respect to each other.} The effect of errors is to create and move pairs of anyons. Measurement of the anyon configuration provides the error syndrome, while error correction corresponds to a pairwise annihilation of anyons which does not form topologically non-trivial loops. {In the limit of large system sizes, such correction is always possible if the error probability is below a certain threshold.} The system can therefore maintain a stable quantum memory even in the presence of a finite anyon density. The Hamiltonian of the toric code model is given by
\be
H_\mathrm{TC} = - \sum_v J_v A_v - \sum_p J_p B_p.
\ee
Normally the case with $J_v=J_p=J>0$ for all $v$ and $p$ is considered. The Hamiltonian $H_\mathrm{TC}$ has the stabiliser space as its ground state and energetically penalises anyon creation. As the $e$ and $m$ anyons are related by a duality transformation we can focus our study on the $e$ anyons without loss of generality.

{\bf Magnetic fields and quantum walks:} Consider the following magnetic field perturbation to the toric code Hamiltonian,
\be
H_h = h \sum_i \sigma^z_i.
\ee
This perturbation commutes with the $B_p$ operators so it has no effect on the $m$ anyons. However, it does not commute with the $A_v$ so it can create, annihilate and transport $e$ anyons. It is instructive to separate this perturbation into two terms $H_h = T + C$ with
\be
T = \sum_n P_n H_h P_n, \,\,\, C = \sum_{n \neq m} P_n H_h P_m.
\ee
Here $P_n$ is the projector onto the subspace of states with $n$ anyons of type $e$, satisfying $\sum_n P_n = \openone$. The operator $T$ transports anyons, while the operator $C$ is responsible for their creation and annihilation and hence is energetically suppressed by the gap. {As such, to the first order of perturbation theory, the effect of $H_h$ can be well described by $T$ alone when $h \ll J$. To the same order, a general field will also act on the $e$ anyons in this way.} The resulting perturbed Hamiltonian describes quantum walks \cite{kempe}, which transport the $e$ anyons around the lattice. {It should be noted that the perturbed Hamiltonian can also be mapped to the 2D Ising model in a transverse field, in which the spin polarized phase corresponds to the toric code.} To proceed let us initially consider only the $e$ anyons, ignoring the $m$'s and the degeneracy associated to them. We then deal with a reduced Hilbert space of dimension $2^{L^2-1}$,  spanned by the states describing the anyonic occupation of each of the $L^2$ vertices. The Hamiltonian of this system can be written as
\bq \label{M}
H_e &=& \sum_{v,v'} M_{v,v'} t_{v,v'} + U \sum_v n_v (n_v-1), \\
M_{v,v'} &=& \delta_{\langle v,v' \rangle} h_i + \delta_{v,v'} J_v,
\eq
where $\delta_{\langle v,v' \rangle}=1$ only when $v$ and $v'$ are neighbouring and $i$ is the corresponding link, and $n_v$ are the number operators for each vertex. The operator $t_{v,v'}$ maps a state with an anyon on the vertex $v$ to one with the anyon moved to $v'$. Furthermore, it annihilates any state without an anyon initially on $v$. {Since the anyons are hardcore bosons, we are interested in the case of $U\rightarrow \infty$.}

If the initial state of the system has no anyons, the Hamiltonian $H_e$ has trivial effect. With no anyons to be moved, no logical errors can occur. Realistically, however, single spin errors will always occur, leading to an ever present population of anyons. When the probability of such errors is low, the resulting anyon configuration will be of sparsely distributed anyon pairs. As such each pair can be considered independently, and so the study of two anyon walks is sufficient to determine the lifetime of the memory. 

The Hamiltonian for the two anyon walks is the projection of $H_e$ onto states of two walkers only. This Hamiltonian is denoted $M^{II}$ and acts upon a $L^2(L^2-1)/2$ dimensional space. This is a significant reduction from the $2^{L^2-1}$ dimensional space of the full Hamiltonian, making numerical studies tractable. The Hamiltonian in Eq. (\ref{M}) accounts only for the $e$ anyons and ignores their non-trivial anyonic braiding with the $m$ anyons that can be initially present in the system. {This should not affect our results, since it has been shown that the effects of the statistics only arise to higher order in perturbation theory than we consider \cite{vidal}. It has also been shown in \cite{gavin} that, for a one dimensional walk of $e$ anyons around a regular pattern of $m$'s, the walk is equivalent to that when no $m$'s are present. We would therefore expect a similar result to apply here, though the random pattern of $m$'s will likely cause the walk to become even slower. }

In the extreme case, where only a single pair of anyons is present a logical error will occur when the anyons walk a distance of order $L/2$ from each other. The quantum walks induced by the field cause this to occur very quickly, in a time linear with the system size, $L$.  A similar argument holds for an anyon density of $\rho=N/L^2$. Here traversing more than a distance of $~\sqrt{\rho}/2$, half the average distance between pairs, will cause error correction to become ambiguous and prone to fail. Hence the dynamic effects of such perturbations destroy fault-tolerance when any non-zero density of anyons is present \cite{kay}, in stark contrast to the case with no field. However, by considering the effects of Anderson localization, fault-tolerance can be regained.
 
{\bf Disorder and localisation:} When the Hamiltonian is disordered, with the $J$ not uniform but taking random values $J_v$ on each vertex, the theory of localisation places an exponential bound on the eigenstates of the single walker Hamiltonian \cite{anderson}. It is expected that this effect persists for the case of two interacting particles \cite{two}, such as the anyon pairs with hardcore repulsion as considered here. We formulate the bound on the eigenstates of the two particle Hamiltonian, $M^{II}$, as follows: When localization takes place, each eigenstate of $M^{II}$ is peaked around a different position state for the anyons $\ket{v_0, v_0'}$. Using this fact, we denote the eigenstates $\ket{E_{v_0,v_0'}}$ according to their peak. The amplitude for any other position state $\ket{v, v'}$ is exponentially suppressed by the distance of $v$ and $v'$ from the peak values. We define this distance by,
\bq
&&d(v, v'; v_0, v_0') = 
\no
&&\,\,\,\,\,\,\,\,\, \min \{d(v;v_0) +d(v';v_0'), d(v;v_0') + d(v';v_0) \},
\eq
where $d(v;v_0')$ is the Euclidean distance between $v$ and $v_0'$, etc. This exponential suppression can then be expressed
\be \label{eig}
|\langle v, v' | E_{v_0, v_0'} \rangle | \leq \exp\left[-\frac{d(v, v'; v_0, v_0')}{2l_{v_0, v_0'}}\right].
\ee
The width of the exponential bound on the eigenstate $\ket{E_{v_0,v_0'}}$ is controlled by the corresponding localisation length, $l_{v_0, v_0'}$. The localization length of the entire Hamiltonian $M^{II}$ is taken to be the maximum over all these.

From the bound on the eigenstates, a bound on the motion of the walkers can be derived. Taking $P(v_f,v_f',v_i,v_i',t)$ to be the probability that the walkers have moved from the vertices $v_i$ and $v_i'$ to $v_f$ and $v_f'$ in a time $t$, we find,
\bq \nonumber \label{bound}
&P&(v_f,v_f',v_i,v_i',t) = |\langle v_f, v_f'|e^{-iM^{II}t}|v_i,v_i' \rangle|^2 \\ \nonumber
&=& |\sum_{v_0,v_0'} e^{-i E_{v_0, v_0'} t} \langle v_f, v_f'|E_{v_0, v_0'}\rangle \langle E_{v_0, v_0'}|v_i,v_i' \rangle|^2 \\ \nonumber
&\leq& \left(\sum_{v_0,v_0'} |\langle v_f, v_f'|E_{v_0, v_0'}\rangle | \, |\langle E_{v_0, v_0'}|v_i,v_i' \rangle|\right)^2 \\
&\leq& \left(\sum_{v_0,v_0'} \exp\left[-\frac{d(v_i, v_i'; v_0, v_0')+d(v_f, v_f'; v_0, v_0')}{2l}\right] \right)^2.
\eq
This sum is dominated by the terms for which $d(v_i, v_i'; v_0, v_0')+d(v_f, v_f'; v_0, v_0')= d$, where $d=d(v_f, v_f'; v_i, v_i')$, since this is the minimum possible value. There are of the order of $d^2$ possible $v_0, v_0'$ that lie directly on the paths between $v_i, v_i'$ and $v_f, v_f'$. Hence the bound can be simplified to $P(v_f,v_f',v_i,v_i',t)\lesssim d^4 \exp[-d/l]$. We may also note that there are of order $d^3$ pairs of vertices $v_f, v_f'$ a distance $d$ from any initial positions $v_i, v_i'$. {The probability $P(d)$ that an anyon pair moves a distance $d$ from its initial position therefore satisfies,}
\be  \label{bound2}
P(d) =O( d^{7} \exp[-d/l] ).
\ee
From this bound the exponential suppression of anyons moving from their initial positions is established.

In order to verify that disorder truly does induce Anderson localization for the two walker Hamiltonian $M^{II}$, and to determine the average value of the localization length for reasonable system sizes, numerical studies have been performed. This is done for a specific case of disorder in which the $J_v$ randomly take values between $J-\gamma/2$ and $J+\gamma/2$ according to a uniform distribution. Here $J$ is the average value of the couplings and $\gamma$ is the strength of the disorder. To calculate the eigenstates of $M^{II}$, the specific values of $J$ and $\gamma$ need not be known. In fact it is only the ratio $\gamma/h$, of disorder strength to magnetic field strength, that is important. {The only constraint on $J$ is that it must be large enough to maintain an energy gap in the full toric code Hamiltonian. It must therefore be at least greater than $\gamma$, but also large enough to overcome the reduction in the gap due to effect of the perturbation $H_h$.}

\begin{figure}[t]
\begin{center}
{\includegraphics[height=4.4cm]{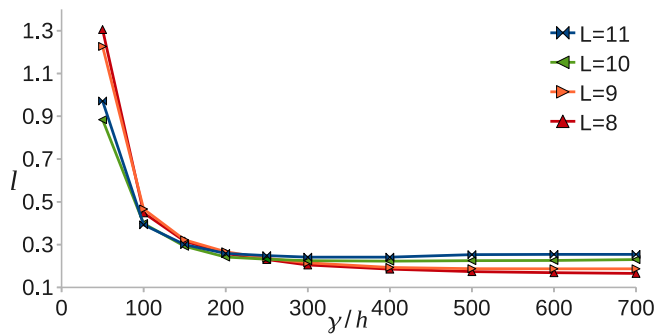}}
\caption{\label{fig1} The average value of the localization length $l$ as the disorder $\gamma/h$ is increased, with each point calculated over 100 samples. The different curves show the results for different system sizes $L$, with the units of $l$ corresponding to the lattice spacing.}
\end{center}
\end{figure}

In our numerical study, once a Hamiltonian $M^{II}$ is generated from the given disorder model, exact diagonalization is used to determine its eigenstates. To calculate $l_{v_0, v_0'}$ for each eigenstate $E_{v_0, v_0'}$, first the peak amplitude and hence the values of $v_0$ and $v_0'$, are found. The probability as a function of the distance $d(v, v'; v_0, v_0')$ is then derived from which $l_{v_0, v_0'}$ can be calculated. The localization length $l$ of the full Hamiltonian $M^{II}$ is then found as the maximum of the $l_{v_0, v_0'}$. {We consider lattice sizes from $L=8$ to $L=11$, for which exponentially localized eigenstates can be clearly seen when $\gamma/h>50$. It is likely that localization also occurs at lower disorder, but it cannot be resolved at these lattice sizes. The resulting values of $l$ for a range of disorder strengths are shown in Fig. \ref{fig1}.}

%In each case $l$ is found to be much  smaller than the separation between lattice sites. Though the value at which $l$ converges as $L\rightarrow \infty$ remains unclear, these results demonstrate that it will be at most $l\sim 1$ for any system size we can reasonably expect to be phsycially realized. The effect of the localization in keeping anyons around their initial positions is therefore very strong.

%\begin{figure}[t]
%\begin{center}
%{\includegraphics[height=4.8cm]{fig1.eps}}
%\caption{\label{fig1} The average value of the localization length $l$ as the system size $L$ is increased, with each point calculated over 100 samples. The different curves show the results for different disorder strengths $\gamma/h$, with the units of $l$ corresponding to the lattice spacing.}
%\end{center}
%\end{figure}

{\bf Localisation and error correction:} The localising effect of the disorder is of great benefit to the toric code quantum memory. Suppression of anyon motion results in suppression of logical errors. Rather than failing in linear time the localisation allows the memory to remain stable in the presence of a finite anyon density, $\rho$, as we now demonstrate. {Consider a regular configuration of anyons distributed as follows. The lattice is partitioned into squares of side length $\lambda$, and a single anyon pair is created in each square. In the absence of anyon motion the memory remains stable, since error correction can be performed by simply by re-annihilating the pairs. When a magnetic field is applied, however, anyons can move out of their squares and the information of the initial pairing is lost. Error correction in this case can be achieved by considering the parity of the number of anyons in each square. Initially, each square has even parity. A single anyon moving between neighbouring squares changes the parity of both. The configuration of odd parity squares may then be used to determine how to best undo the errors caused by the anyon motion. This is possible so long as the probability of an anyon moving over each boundary between neighbouring squares, $p$, is less than the critical value $p_c \approx 0.11$ \cite{kitaev}.} For the case of no disorder, the anyons move quickly and $p$ will exceed $p_c$ in time linear with $\lambda$. In the presence of disorder each anyon pair will obey the bound of Eq. (\ref{bound2}). As such, there will exist a square size $\lambda_c$ for which $p$ is kept below $p_c$, and hence the memory will remain stable, for arbitrarily long times. This gives a critical density of $\rho_c=\lambda_c^{-2}$. Similarly, in the case that anyon pairs are created by random spin errors, we may expect the standard error correction procedure to succeed reliably when the average distance between pairs is at least $\lambda_c$, leading to the same critical density.

\begin{figure}[t]
\begin{center}
{\includegraphics[height=5cm]{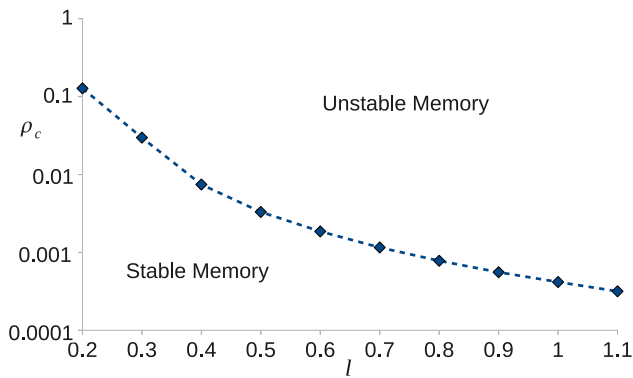}}
\caption{\label{fig2} {The behavior and approximate values of the critical anyon density $\rho_c$ as a function of the localisation length, $l$. This is calculated by numerically solving for the side length $\lambda_c$ at which the probability $p$ of anyons moving between squares reaches the critical value $p_c$ where error correction fails.}}
\end{center}
\end{figure}

{It is important to determine how the critical side length $\lambda_c$ and the corresponding critical density $\rho_c$ behave as a function of the localisation length, $l$. To do this, first note that the probability of an anyon moving to a neighbouring square requires $d$ to be at least half the square size, $\lambda/2$, giving  $p< \sum_{d>\lambda/2}P(d)$. Using Eq. (\ref{bound2}) it can then be deduced that $p = O( (\lambda/2)^{7} \exp[-\lambda/(2l)] )$.  The critical side length $\lambda_c$ is that for which $p=p_c$. This can be approximated by the solution to $(\lambda_c/2)^{7} \exp[-\lambda_c/(2l)]=p_c$, since the coefficient on the asymptotic behaviour of $p$ will not have a significant effect.  The critical density is then calculated according to $\rho_c=\lambda_c^{-2}$. The results of the numerical solution to this problem are shown in Fig. \ref{fig2}. These demonstrate the behavior $\rho_c$ as a function of $l$ and give an estimate for its order of magnitude. It is found that $\lambda_c = O(l)$ and $\rho_c = O(l^{-2})$, as one might expect from the exponential factor of Eq. (\ref{bound2}).}

{\bf Conclusions:} Magnetic field perturbations on the toric code induce quantum walks of anyons, which quickly destroy any stored information when anyons are present. However, we have shown that disorder induces exponential localization which suppresses the anyon motion. This allows the memory to remain stable even when a finite anyon density is present. Since disorder will be inherent in any experimental realisation of topological systems, e.g. with Josephson junctions \cite{Doucot}, the effect described here is expected to play a significant role in their behaviour. Localization will also protect against Hamiltonian perturbations in other topological models, including those of non-Abelian anyons. The prospect of purposefully engineering disorder into topological systems to benefit from further localization effects, for both coherent and incoherent errors, is a subject of continuing study.

%{\bf Conclusions:} Here we have demonstrated that magnetic field perturbations on the toric code induce quantum walks of anyons, which quickly destroy any stored information when anyons are present. When the field is sufficiently weak in comparison to disorder in the couplings of the Hamiltonian, exponential localization is induced that suppresses the anyon motion. This allows the memory to remain stable even when a finite anyon density is present. Since disorder will be inherent in any experimental realisation of topological systems, e.g. with Josephson junctions \cite{Doucot}, the effect described here is expected to play a significant role in the behaviour of these systems. Anderson localization will also play a role in non-Abelian models, where the anyons are used to store and process quantum information. By preventing these anyons from being moved in error by Hamiltonian perturbations, localization preserves the memory and prevents logical errors by unwanted braidings. This work suggests the prospect of purposefully engineering disorder into topological systems to benefit from further localization effects, for both coherent and incoherent errors, which is a subject of continuing study.

{\bf Additional Note:} Complementary results have been obtained independently by Cyril Stark, Atac Imamo\u{g}lu and Renato Renner \cite{stark}.

{\bf Acknowledgements:} We would like to thank Roberto Alamino and Alioscia Hamma for inspiring conversations, Alastair Kay for critical reading of the manuscript and Robert Heath for working with us on related issues. This work was supported by EPSRC and the Royal Society.

\end{document}